\documentclass[twocolumn]{article}
\usepackage{graphicx}
\usepackage{dcolumn}
\usepackage{bm}

\usepackage[utf8]{inputenc}
\usepackage[T1]{fontenc}
\usepackage{mathptmx}
\usepackage{amsmath}
\usepackage{algorithm}
\usepackage{enumerate}
\usepackage{algorithm}
\usepackage{algpseudocode} 
\usepackage[table,xcdraw]{xcolor}
\setkeys{Gin}{width=\linewidth}
\usepackage[mode=buildnew]{standalone}
\usepackage{placeins}
\usepackage{tikz}
\usepackage{hyperref}
\usepackage[toc,page]{appendix}
\usepackage[mode=buildnew]{standalone}

\usetikzlibrary{positioning, shapes.geometric, arrows}

\usetikzlibrary{shapes.geometric, arrows, positioning, fit}
\tikzstyle{startstop} = [rectangle, rounded corners, minimum width=3cm, minimum height=1cm,text centered, draw=black]
\tikzstyle{io} = [trapezium, trapezium left angle=70, trapezium right angle=110, minimum width=3cm, minimum height=1cm, text centered, draw=black]
\tikzstyle{process} = [rectangle, minimum width=3cm, minimum height=1cm, text centered, text width=3cm, draw=black]
\tikzstyle{decision} = [diamond, minimum width=3cm, minimum height=1cm, text centered, draw=black]
\tikzset{main node/.style={circle,,draw,minimum size=1cm,inner sep=0pt} }
\tikzstyle{arrow} = [thick,->,>=stealth]

\usepackage[backend=biber, 
                url = false,
                giveninits=true]{biblatex}
\begin{document}

\title{Don't go chasing artificial waterfalls:\\
Simulating cascading failures in the power grid and the impact of artificial line-limit methods on results}
\author{J. Bourne \thanks{ucabbou@ucl.ac.uk} \thanks{University College London, Gower Street, WC1E 6BT, UK} 
\and A. O'Sullivan \footnotemark[2] \and E. Arcaute \footnotemark[2]}

\date{\today}

\maketitle

\begin{abstract}
Research into cascading failures in power-transmission networks requires detailed data on the capacity of individual transmission lines. However, these data are often unavailable to researchers. As a result, line limits are often modelled by assuming they are proportional to some average load. Little research exists, however, to support this assumption as being realistic. In this paper, we analyse the proportional-loading (PL) approach and compare it to two linear models that use voltage and initial power flow as variables. 
In conducting this modelling, we test the ability of artificial line limits to model true line limits, the damage done during an attack and the order in which edges are lost. we also test how accurately these methods rank the relative performance of different attack strategies.
We find that the linear models are the top-performing method or close to the top in all tests. In comparison, the tolerance value that produces the best PL limits changes depending on the test. The PL approach was a particularly poor fit when the line tolerance was less than two, which is the most commonly used value range in cascading-failure research. We also find indications that the accuracy of modelling line limits does not indicate how well a model will represent grid collapse. In addition, we find evidence that the network's topology can be used to estimate the system's true mean loading.
The findings of this paper provide an understanding of the weaknesses of the PL approach and offer an alternative method of line-limit modelling.
\end{abstract}

\section{Introduction}

The networked structure of power-transmission grids makes them susceptible to cascading failures. A cascading failure occurs when a single failure or a small number of failures propagate through a system, wreaking havoc. While major blackouts caused by cascading failures are rare, they affect very large numbers of people and have significant financial consequences. A major blackout in 2003 in the US Northeast is estimated to have cost around USD6 billion \cite{walker_counting_2014}. In 2012 a cascading failure in the Indian power grid caused a loss of power to 600 million people \cite{guo_critical_2017}. More recently in 2019 \cite{BBC_Uru_Arg}, a failure in the interconnector between Argentina and Uruguay caused a blackout that affected nearly the entirety of both countries, close to 50 million people. Such massive failures occur through the cascading effect, and their distribution follows a power law \cite{dobson_complex_2004, carreras_north_2016, hines_large_2009}. This means there are a large number of blackouts of little consequence and a very small number that are of the scale described in the examples given previously. This paper explores a commonly used method to simulate cascading failures in the power grid, known as proportional loading, and finds that inaccurate estimation of the system tolerance produces results that do not represent the true behaviour of the grid.

Given the potential magnitude of cascading failures, it is unsurprising that researchers are looking for ways to reduce their impact and frequency. One method of understanding cascading failures is through network science. The cascading failures are stimulated using targeted attacks on network nodes or edges. Substantial work in this regard has focused on developing `vulnerability metrics' or `attack strategies' that identify the order in which nodes should be attacked to cause maximum damage to the power grid.

When researchers first started analysing power grids using network science, the techniques applied used purely topological information about the power-grid structure \cite{cuadra_critical_2015}. As research developed, the power-grid's electrical properties were incorporated into the analysis, creating the `Extended topology' \cite{bompard_analysis_2009}. Specifically, the extended topology integrates the power flowing in the network with the topological features. Electricity is transmitted through the power grid using alternating current (AC) to reduce power loss over long distances. However, as the AC power-flow equations are challenging to solve, researchers often use direct-current (DC) flow equations as an approximation. A recent literature review showed that 81\% of studies that involved power flow used the DC approximations \cite{cuadra_critical_2015}.
we use the DC flow calculations $\mathbf{f}=\mathbf{CA(A^TCA)^{-1}p}$ described by \cite{pepyne_topology_2007} and \cite{arianos_power_2009}. In this equation, $\mathbf{A}$ is the adjacency matrix with the slack bus removed to make the system invertible, $\mathbf{C}$ is a diagonal matrix of the line susceptance and $\mathbf{p}$ is the power injected at each line.

Many studies create synthetic networks \cite{pepyne_topology_2007, arianos_power_2009, wu_study_2017, hines_centrality_2008} or use the topological structure of a real power grid without the line limits \cite{yan_cascading_2015, wu_study_2017, buldyrev_catastrophic_2010, kinney_modeling_2005, schafer_dynamically_2018}. Knowing these limits is necessary to detect when a line has tripped in a simulation. 
Although some open-data solutions are being developed \cite{medjroubi_open_2017}, the lack of datasets with line limits remains a problem.
 Little modelling or simulation work has been done using real line limits \cite{hines_topological_2010-1} for this reason. If these limits are not available in the data, an estimate needs to be made. A common technique for this is to use proportional loading (PL). When using the PL approach, the line limits are set at a fixed proportion of the amount of power flowing in each line at initiation \cite{motter_cascade-based_2002, kinney_modeling_2005, pepyne_topology_2007, yan_cascading_2015, zhu_revealing_2014, koc_impact_2014, ouyang_comparisons_2014}. The PL of networks is usually defined as $f_i^\text{max} = \alpha \left | f_{i}^c \right |$, where  $\left | f_i^c \right |$ is the absolute power flow for line $i$ under initial conditions, $f_i^\text{max}$ is the line limit and $\alpha$ is the tolerance factor. The PL approach makes strong assumptions about power-grid design for which there is no supporting evidence.

As there is no direct comparison between PL and real line limits, it is difficult to know how accurately PL and the simulations based on it reflect real-grid behaviour or whether a more realistic method can be created. In this paper, we address this gap. We have access to a dataset of the UK high-voltage power grid that includes the generation and load nodes with capacities in MW, as well as the line limits in MW. 

We compare the real line limits of the network against the PL values of $\alpha$ between 1 and 50, the results produced by two linear models and a topological analysis. We simulate random attacks on the power grid and analyse how well the artificial line limits model the damage caused by the attack, as well as the order in which the edges are lost due to cascading. We also measure how well each line-limit method ranks the relative effectiveness of different grid-attack strategies. The findings of this paper provide an understanding of the PL approach's weaknesses and provide an alternative method of line-limit modelling.

\section{Method}
The analysis was conducted in four stages to test different aspects of how well the artificial line limits model the real line limits. The code used to generate the results in this paper is available on GitHub \cite{bourne_powergridnetworking_2018, bourne_proportionalloading_2018}.

\begin{enumerate}
    \item We look at the network's real line-limit distribution. We then compare the accuracy of PL and the modelled limits against the real line limits.

    \item We simulate random attacks on the grid using the DC-flow approximations. We attack the grid until it collapses completely, repeating the process 100 times. We compare the mean damage and standard deviation of the artificial line limits and the real line limits at each stage of the attacks.  

    \item We compare the rank order in which edges are lost due to cascades between the artificial and real line limits. We then calculate the correlation coefficient, which shows us which artificial line-limit method most accurately represents the behaviour of grid collapse.
    
    \item To test whether the artificial line limits can compare vulnerability metrics, 
    we attack the grid using five different strategies. We measure the ability of the artificial line limits to accurately represent the true ranking of each strategy using the real line limits.
\end{enumerate}

The dataset we use in this paper describes the physical and electrical structure of the UK national grid. The dataset contains 512 nodes representing substations around the UK. The nodes are connected with 698 transmission lines that are rated 132kV, 275kV and 400kV. The network has a mean degree of 2.73, and the average unweighted nodal distance is 11.7. The assortativity, clustering and mean normalised centrality are close to zero (-0.06 0.1 and 0.03, respectively). The dataset includes all the information required to perform the DC load-flow calculations. This information includes line-node connections, line-reactance node load and node generation. Unusually for a power-grid dataset, it also provides line limits, making it possible to test the PL approach. 

This paper defines the simulation using five parameters: physics model, element, attack type, removal method and load profile. The physics model used in this analysis is DC flow or topological. That is the power model uses the DC power flow assumptions, or is simply a topological analysis.
The elements attacked are the nodes. The attack type is `fixed', meaning that the order in which the nodes will be removed is generated before the attack begins. The order in which nodes are removed is sequential. This means that only a single node is targeted for removal each round. However, other nodes may be lost due to a cascading failure. A summary of the parameters is shown in Table \ref{tab:PEARL} The last parameter is load profile. In this analysis, we use a single load profile based on the year-round baseload provided as part of the dataset. These simulation parameters form the acronym PEARL. (The PEARL framework is discussed in greater depth in Appendix \ref{sect:PEARL}).

    \begin{table}
\centering
\caption{PEARL settings used in this paper.}
\label{tab:PEARL}
\begin{tabular}{l|l}
Class        & Types                     \\ \hline
\textbf{P}hysics       & Cascading DC, Topological \\
\textbf{E}lement      & Node\\
\textbf{A}ttack Type  & Fixed \\
\textbf{R}emoval      & Sequential \\
\textbf{L}oad Profile & Single         
\end{tabular}
\end{table}

\subsection{Artificial line limits and real line limits}
To understand how artificial line limits differ from real line limits, we use 13 $\alpha$ values for PL, the topological approach and the line limits generated from two linear models. 
we use $\alpha$ values of 1.05, 1.1, 1.2, 1.3, 1.5, 2, 3, 5, 7, 10, 15, 20 and 50. The $\alpha$ values were chosen based on values used in other papers, typically 1 to 5, and extended until 50. The topological analysis is identical to an $\alpha$ level of infinity.

Line limits are caused by three main factors: thermal-limits in the cable, voltage drop and frequency stability. These three factors are all related to the line length and its voltage \cite{dunlop_analytical_1979}. To keep the models similar to PL, 
we create a model that uses only the initial power flow on the line and a second model that uses initial power flow and the voltage as voltage data are available in several open datasets \cite{entsoe-e_grid_nodate, rivera_opengridmap:_2015,contributors_planet_2017}. The Voltage Power Flow (Volt PF) model has the form $y_i = \beta_0 1 + \beta_{f} x_{if} + \beta_{v275} x_{iv275}+ \beta_{v400} x_{iv400}$, where $y_i$ is the line limit of the $i$th power line, $x_{if}$ is the initial power flow of the $i$th power line in MW and $x_{iv275}, \; x_{iv400}$ are binary vectors representing the voltage level of line $i$. The coefficients are $\beta_f, \; \beta_{v275}, \; \beta_{v400}$, respectively, while $\beta_0$ is the model bias. The power-flow-only model (PF model) has the form  $y_i = \beta_0 1 + \beta_{f} x_{if}$.
The models are trained using tenfold cross-validation. These models will be compared to the PL approach throughout the paper. Ideally, we could use the models to predict the line limits of a separate network. However, we only have access to the real line limits of a single network. As this is the case, we will use the line limits predicted in each of the validation sets as the predicted line limits for the attack simulation. The artificial line limits are compared to the real line limits using $R^2$, Root Mean Square Error (RMSE) and Mean Average Percent Error (MAPE). 

\subsection{Comparing attack damage between the PL approach and real line limits}
\begin{figure}
\resizebox{\linewidth}{!}{%
    \centering
    \begin{tikzpicture}[node distance=2cm]

\node (start) [startstop] {Remove Node};
\node (PowCalc) [process, below of=start, yshift = -1cm] {Calculate Power Flow};

\node (OverLim) [decision, right of=PowCalc, xshift = 2cm] {Over limit?};

\node (Remove) [process, below of=OverLim, yshift=-1cm] {Remove Lines};

\node (Rebalance) [process, below of=PowCalc, yshift=-1cm] {Rebalance Grid Power};

\node (stopconds) [decision, right of = OverLim, xshift = 2.5cm, text width=2cm] {Are stop conditions met?};


\node (stop) [startstop, below of = stopconds, yshift = -1cm] {Attack Finished};

\draw [arrow] (start) -- (PowCalc);
\draw [arrow] (PowCalc) -- (OverLim);
\draw [arrow] (OverLim) -- node[anchor=east] {yes}(Remove);
\draw [arrow] (Remove) -- (Rebalance);
\draw [arrow] (Rebalance) -- (PowCalc);

\draw [arrow] (OverLim) -- node[anchor=south] {no}(stopconds);


\draw [arrow,rounded corners=5pt] (stopconds.north) |- node[anchor=south] {no}(start.east);
\draw [arrow] (stopconds) -- node[anchor=east] {yes}(stop);

\node[draw, red,thick,dashed,inner sep=2mm,label=below:Cascade,fit=(PowCalc) (OverLim) (Rebalance) (Remove)] {};

\node[draw, inner sep=7mm,label=below:Attack the grid,fit=(stop) (start)] {};

\end{tikzpicture}
}
    \caption{Schematic of the node removal process}
    \label{fig:AlgoSchem}
\end{figure}
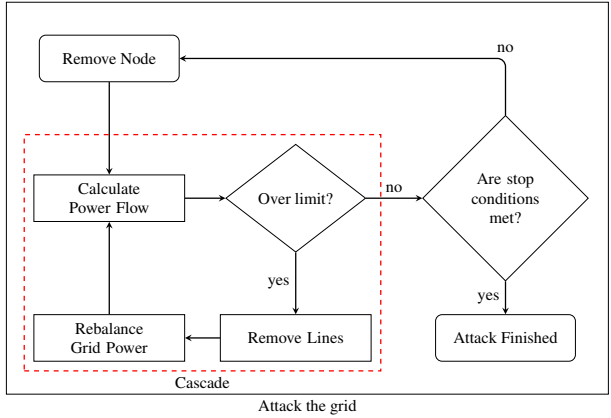

we simulate the grid under attack by assigning a random-node attack rank to each network node. We then remove the node with Rank 1. As node removal can cause overloading, we recalculate the power flow and remove any lines that exceed their maximum line limit. We then rebalance the power and load in the network. Line removal can cause the network to break into subcomponents, so we remove any nodes that are in a subcomponent that has no power. We then recalculate power flow. This process continues until no further removals occur. We then find the node with the next lowest attack rank and remove it, repeating the process until no nodes remain in the network. The process is described in Algorithm \ref{algo:attackthegrid} for the graph $G$, the set of $n$ nodes/vertices $\mathcal{V}$ and the set of $m$ edges $E$. A schematic of the process is shown in Figure \ref{fig:AlgoSchem}. 

\begin{algorithm}[H]
\caption{Attack the grid}\label{algo:attackthegrid}
\begin{algorithmic}[1]
\Procedure {AttackTheGrid}{$G$}

\State $\mathcal{V} \leftarrow  V(G)$
\State $E \leftarrow  E(G)$

\While{$\mathcal{V} \neq \emptyset$ }
\State remove $\text{min}_i(v_i)$ 
\Comment $i$ is the order in which nodes will be attacked
\Repeat
\State calculate power flow in $G$
\For{$e \in E$}
\State  remove $e$ if $f_e > f_e^{max}$ \Comment Power flow exceeds line limit
\EndFor
\State  rebalance generation and supply
\For{$v \in \mathcal{V}$}
\State  remove $v$ if subgraph has no power
\EndFor
\Until{ $f_e < f_e^{max}$ for all $e$}
\EndWhile
\EndProcedure
\end{algorithmic}
\end{algorithm}

During the attack simulation, two different damage metrics will be used to analyse attack progression. These are giant-component size, measured as the largest connected component, and blackout size, measured as total MW lost. These metrics will be compared to the graph's original state using $\Delta P_x= 1-\frac{P_1-P_x}{P_1}$, where $P_1$ is the complete graph and $P_x$ is the graph after attack $x$. This way of measuring damage returns a percentage between 0 (no damage) and 100 (complete grid collapse).

We note here that using the largest connected component as a measure might not be representative of the physical processes that occur on the power grid. For example, say we have a network that has 100 nodes but only two generators, where the first generator produces 99\% of the electricity and the other generates the remaining 1\%. If the big generator fails, the giant-component size is still 99\%, indicating that the network is almost unaffected even though there only is 1\% of the necessary power. Despite this drawback, the metric's simplicity has made it a popular choice \cite{andrea_pagani_power_2014, motter_cascade-based_2002, rosas-casals_topological_2007,pagani_towards_2011}, and so it will be included here.
Blackout size is a metric from the extended topology. It measures the loss of system power in MW. This metric is popular and has several different names with slightly different implementations that produce similar results. It has been called `loss of load' \cite{wang_electrical_2011}, blackout size \cite{hines_topological_2010-1}, `load shed' \cite{carreras_north_2016}, `power supply' \cite{ouyang_comparisons_2014} and `total loss of power' \cite{brancucci_martinez-anido_european_2012} among other names. This metric does not suffer the problem described above regarding the size of the giant component.

\subsection{Comparing the order in which edges are lost due to cascade}
Accurately modelling the damage done in an attack provides much insight into a grid's vulnerability. Sometimes, however, it is also useful to know the order in which nodes or edges were lost as this information is used in some vulnerability metrics \cite{yang_vulnerability_2017, hines_cascading_2017, zhu_revealing_2014}. One caveat here, though, is that if the order of the nodes being lost differs depending on the line limits, the results of such analyses will not be reliable. To explore the robustness of loss order in relation to line limits, we correlate the order in which nodes are lost during an attack for artificial line limits with real line limits.

Tables \ref{table:SimAttackOrder} and \ref{table:NodeLossOrder} give a toy example of how we compare node-loss order. First, $k$ node removal orders are generated. Then all $k$ simulations are run using each line-limit type. Table \ref{table:NodeLossOrder} shows Simulation 1 for the real line limits and a line limit of $\alpha = 3$. In the table, the yellow cells indicate a node that has been targeted for removal, while the other nodes were lost due to the cascade. We find the similarity of network collapse by correlating the round lost to the cascade of the nodes. This means we exclude nodes that were targeted for removal. In the example, that means only nodes F, G and H can be compared.

For each of the $k$ simulations, we use the Spearman's correlation $\rho_k = \frac{\text{cov}(\text{rg}_{x,k},\text{rg}_{y,k})}{\sigma_{\text{rg}_{x,k}}\sigma_{\text{rg}_{y,k}}}$, where $\text{rg}_{x,k}$ and $\text{rg}_{y,k}$ are the rank of $x_k$ and $y_k$, respectively, for the $n$ nodes in the network. In this case, $x_k$ is the vector of the node-removal rounds of the artificial line limit for Simulation $k$ and $y_k$ is the vector of the node-removal round for the real line limits for Simulation $k$. In the example, this means $x_1 = {2,4,4}$ and $y_1 = {1,3,2}$, giving a correlation of 0.866. 

For the real experiment, we generate 100 attack orders for the 512 nodes, producing 100 correlation scores per line-limit method. We can then see how similar the artificial line limits' collapse order is to that of the real line limits.

\begin{table}
\centering
\label{my-label}
\caption{The node removal orders are randomly generated k times.}
\begin{tabular}{l|llll}
\begin{tabular}[c]{@{}l@{}}Node\\ ID\end{tabular} & Sim 1 & Sim 2 & $\cdots$ & Sim k \\
\hline \\
A       & 1     & 4     & $\cdots$ & 3     \\
B       & 2     & 2     & $\cdots$ & 8     \\
C       & 3     & 1     & $\cdots$ & 7     \\
D       & 4     & 6     & $\cdots$ & 6     \\
E       & 5     & 5     & $\cdots$ & 1     \\
F       & 6     & 7     & $\cdots$ & 2     \\
G       & 7     & 8     & $\cdots$ & 5     \\
H       & 8     & 3     & $\cdots$ & 4    
\end{tabular}
\label{table:SimAttackOrder}
\end{table}

\begin{table}
\centering
\caption{The numbers indicate the round in which the node was lost in Simulation 1 for the real limits and $\alpha = 3$. Yellow nodes were targeted for removal and so are not included when calculating removal similarity.}
\begin{tabular}{l|ll}
\begin{tabular}[c]{@{}l@{}}Node\\ ID\end{tabular} & \begin{tabular}[c]{@{}l@{}}Real\\ Limits\end{tabular} & $\alpha = 3$              \\
\hline \\
A                                                   & \cellcolor[HTML]{FCFF2F}1                             & \cellcolor[HTML]{FCFF2F}1 \\
B                                                   & \cellcolor[HTML]{FCFF2F}2                             & 1 \\
C                                                  & 1                                                     & \cellcolor[HTML]{FCFF2F}2 \\
D                                                  & \cellcolor[HTML]{FCFF2F}3                             & \cellcolor[HTML]{FCFF2F}3 \\
E                                                  & \cellcolor[HTML]{FCFF2F}4                             & 1                         \\
F                                                  & 2                                                     & 1                         \\
G                                                   & 4                                                     & 3                         \\
H                                                     & 4                                                     & 2                        
\end{tabular}
\label{table:NodeLossOrder}
\end{table}

\subsection{Comparing vulnerability ranking accuracy}

The choice of attack strategy can have a substantial impact on results. Different strategies will damage the network at different rates throughout the attack. When comparing attack strategies, it may be that only the relative performance is important. In this analysis, we compare the ability of different line-limit methods to accurately rank attack strategies. We compare five different attack strategies across eight $\alpha$ values, the modelled limits and the topological limits. The $\alpha$ values used in this analysis are 1.05, 1.1,1.5, 2, 5,  10, 20 and 50.

The attack strategies we use in this analysis are a mixture of topological methods and extended topology. The strategies are entropic degree \cite{bompard_analysis_2009} using line limit, entropic degree \cite{bompard_analysis_2009} using initial power flow, degree, centrality and electrical centrality \cite{wang_electrical_2011}.

Using each attack strategy to define the node-removal order, we simulate an attack until we achieve complete grid collapse. After each attack round (node removed), we rank the strategies by total blackout size. The strategy that caused the most damage to the network gets Rank 1 and the strategy that caused the least damage gets Rank 5. We then compare the rankings obtained by each artificial limit with the real line limits. Using RMSE, we evaluate which line-limit method has the lowest error relative to the real line limits.

\section{Results}
The results are split into two sections. The first section describes the results of the linear model used to generate line limits, and the second section compares the performance of the different artificial line limits to the real line limits

\subsection{Modelling line limits}
\begin{figure}
    \centering
    \includegraphics{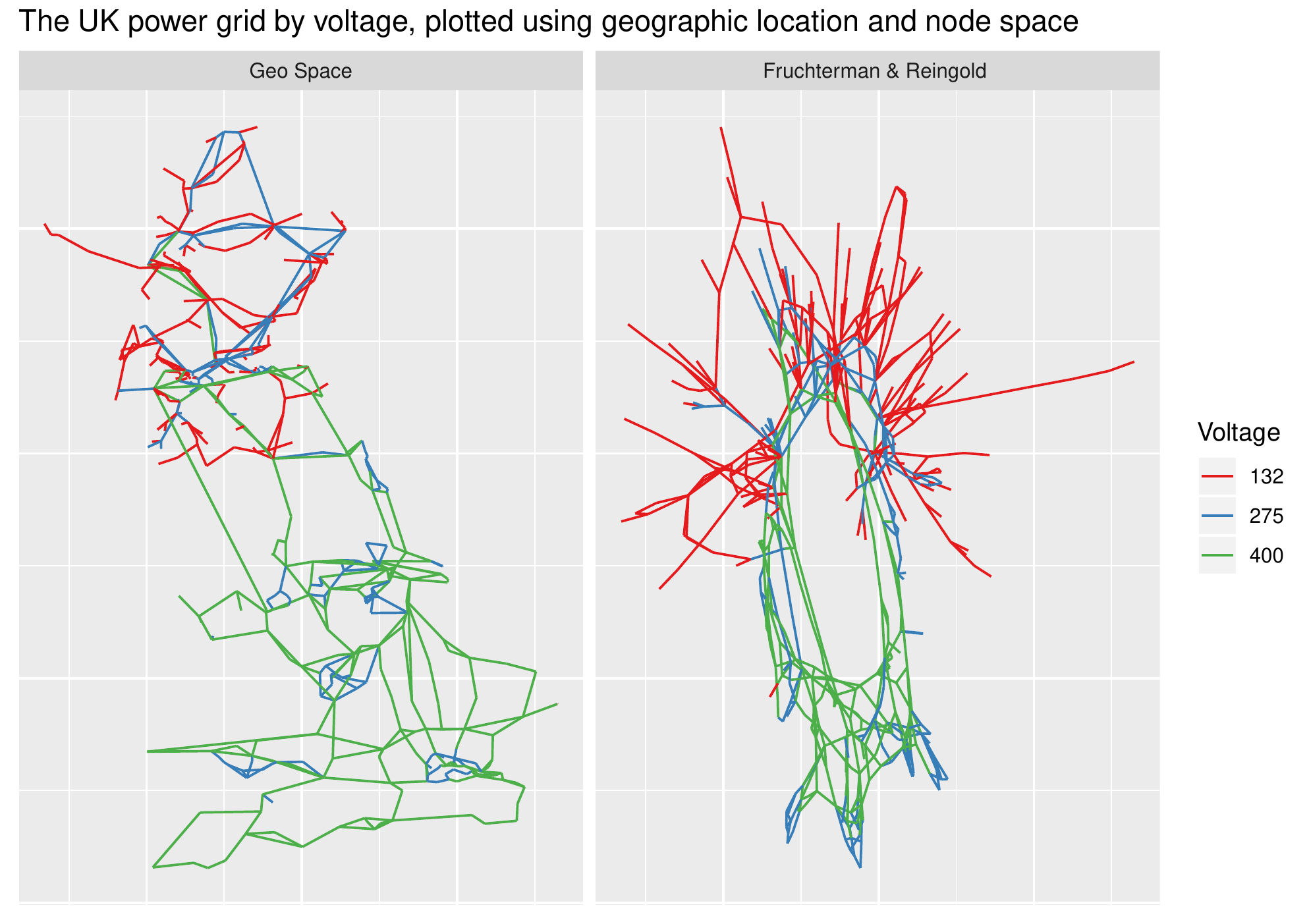}
    \caption{UK power-grid voltage level shown in geographical space and graph space. The 132-volt sections of the network are much less densely connected than the 275 and 400-volt sections}
    \label{fig:VoltageMap}
\end{figure}

We first create a linear model that predicts line limits. The variables used are voltage and initial power flow. The network is represented in Figure \ref{fig:VoltageMap} using a UK geographical location and a graph space representation using force expansion.
As we only have a single network, we use tenfold cross-validation to train on 90\% of the data and then predict using the remaining 10\%. The model coefficients are all quite stable with a proportionally small standard deviation.The coefficients were all positive across all folds. The Volt PF model shows that as voltage increases so does line limit. This is intuitive as high-voltage cables are often used for bulk-power transmission (see Table \ref{tab:ModCoeffs}).
\begin{table}[ht]
\centering
\caption{Mean model coefficients from 10-fold cross-validation} 
\begin{tabular}{rlrr}
  \hline
 & Coefficients & PF \& Voltage Model & PF Model \\ 
  \hline
1 & Intercept & 2.30 & 2.71 \\ 
  2 & PF (in 1000s) & 2.46 & 7.57 \\ 
  3 & Voltage 275 & 0.68 &  \\ 
  4 & Voltage 400 & 1.00 &  \\ 
   \hline
\end{tabular}
\label{tab:ModCoeffs}
\end{table}

\subsection{Comparing performance of different line limits}

\begin{figure}
    \centering
    \includegraphics{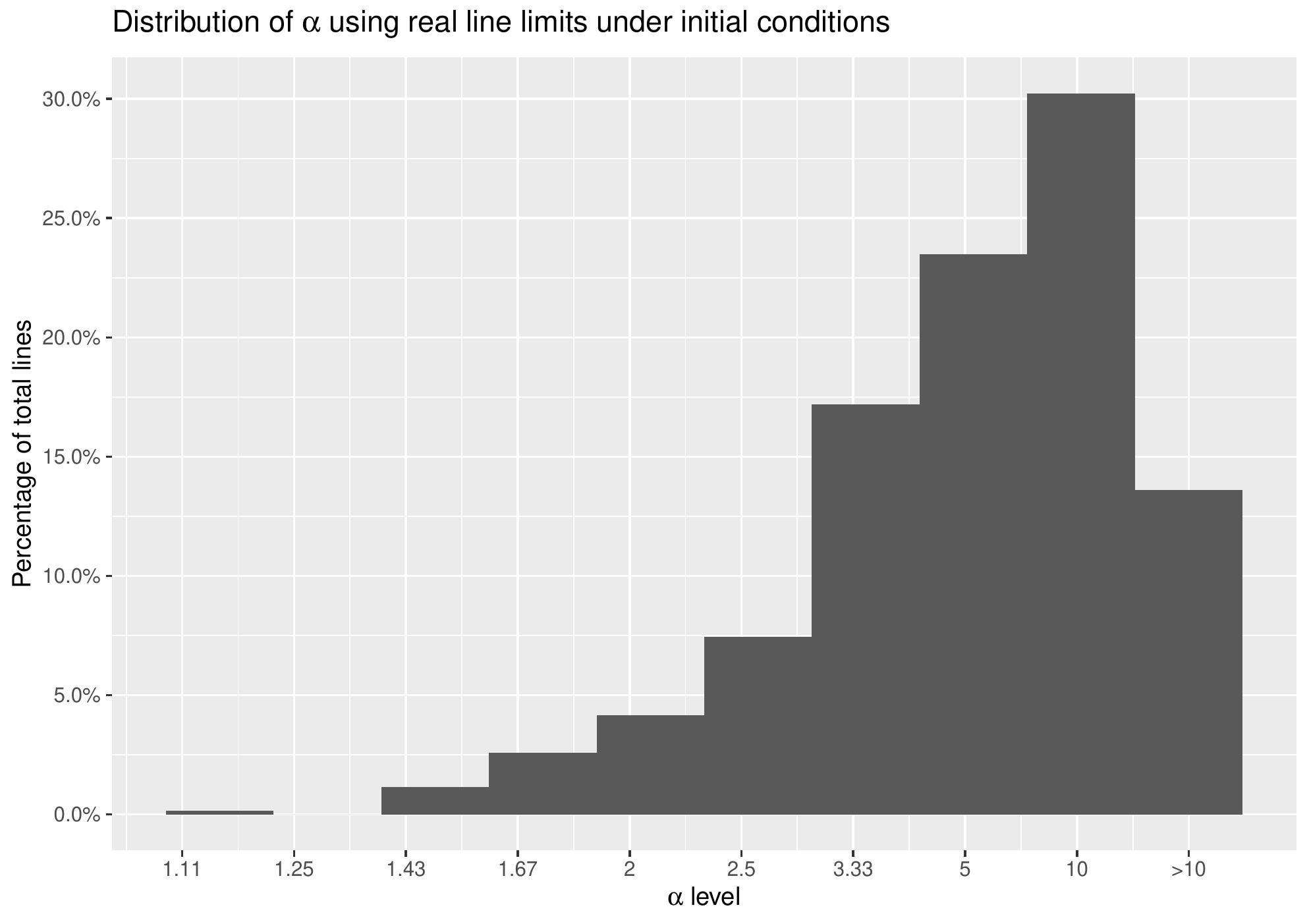}
    \caption{The distribution of the loading is left skewed with a mean $\alpha$ of 5.12.}
    \label{fig:LoadLevel}
\end{figure}

An inspection of the power-grid tolerance distribution under initial conditions shows that the system is not proportionally loaded. Figure \ref{fig:LoadLevel} shows the $\alpha$ distribution of the power grid under initial loading. The power grid has a mean $\alpha = 5.12$ and a median $\alpha = 6.12$ These tolerances are much higher than those used in the literature, which are typically less than two.

Correlating PL limits with real limits shows an $R^2$ of 0.5. Using a range of $\alpha$ values, we find that the minimum MAPE and RMSE are 0.52 and 1380 with $\alpha$ values of 3.6 and 3.5, respectively. The Volt PF model greatly outperforms all PL models and the PF model (see Table \ref{tab:ModError}). The PF model has very poor performance across all three metrics, which shows that it is not an effective method to represent line limits accurately.

\begin{table}[ht]
\centering
\caption{Accuracy of modelling line-limits} 
\begin{tabular}{rlrrr}
  \hline
 & Model & Rsq & RMSE & MAPE \\ 
  \hline
1 & Volt Power-Flow & 0.64 & 1037.14 & 0.43 \\ 
  2 & Power-Flow & 0.13 & 4446.96 & 1.07 \\ 
   \hline
\end{tabular}
\label{tab:ModError}
\end{table}

\begin{figure}
    \centering
    \includegraphics{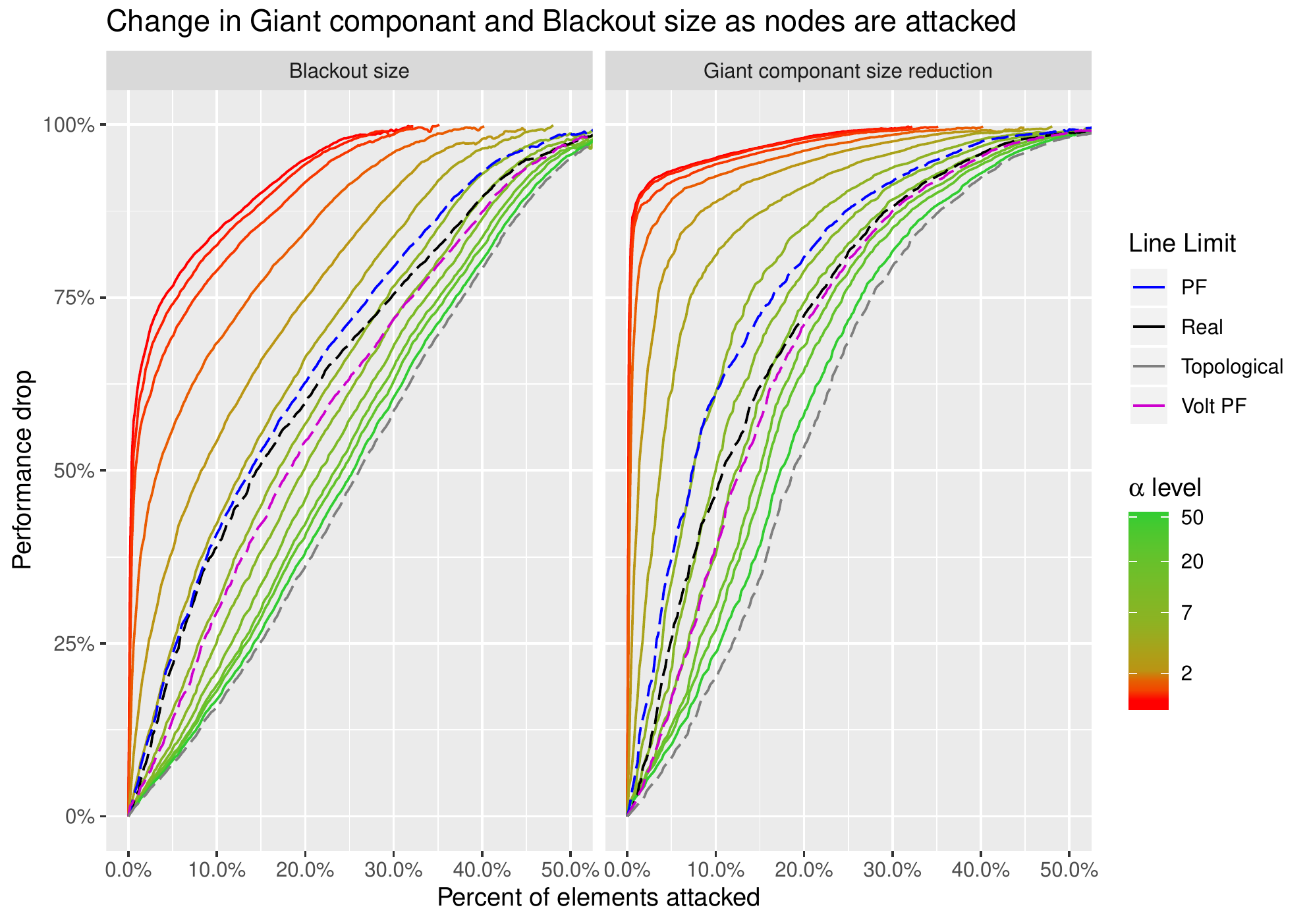}
    \caption{Mean damage done in terms of giant component and blackout size across 100 simulations, across all 13 $\alpha$ levels and the linear models. There is a log relationship between $\alpha$ and damage done for a given number of elements attacked.}
    \label{fig:DamageDone}
\end{figure}

We explore the damage done during random attack across different values of alpha. We see that that alpha has a logarithmic relationship with the damage for a given number of attacks. Figure \ref{fig:DamageDone} shows that for both damage metrics the $\alpha$ levels only converge at grid collapse and do not cross each other before that.

\begin{figure}
    \centering
    \includegraphics{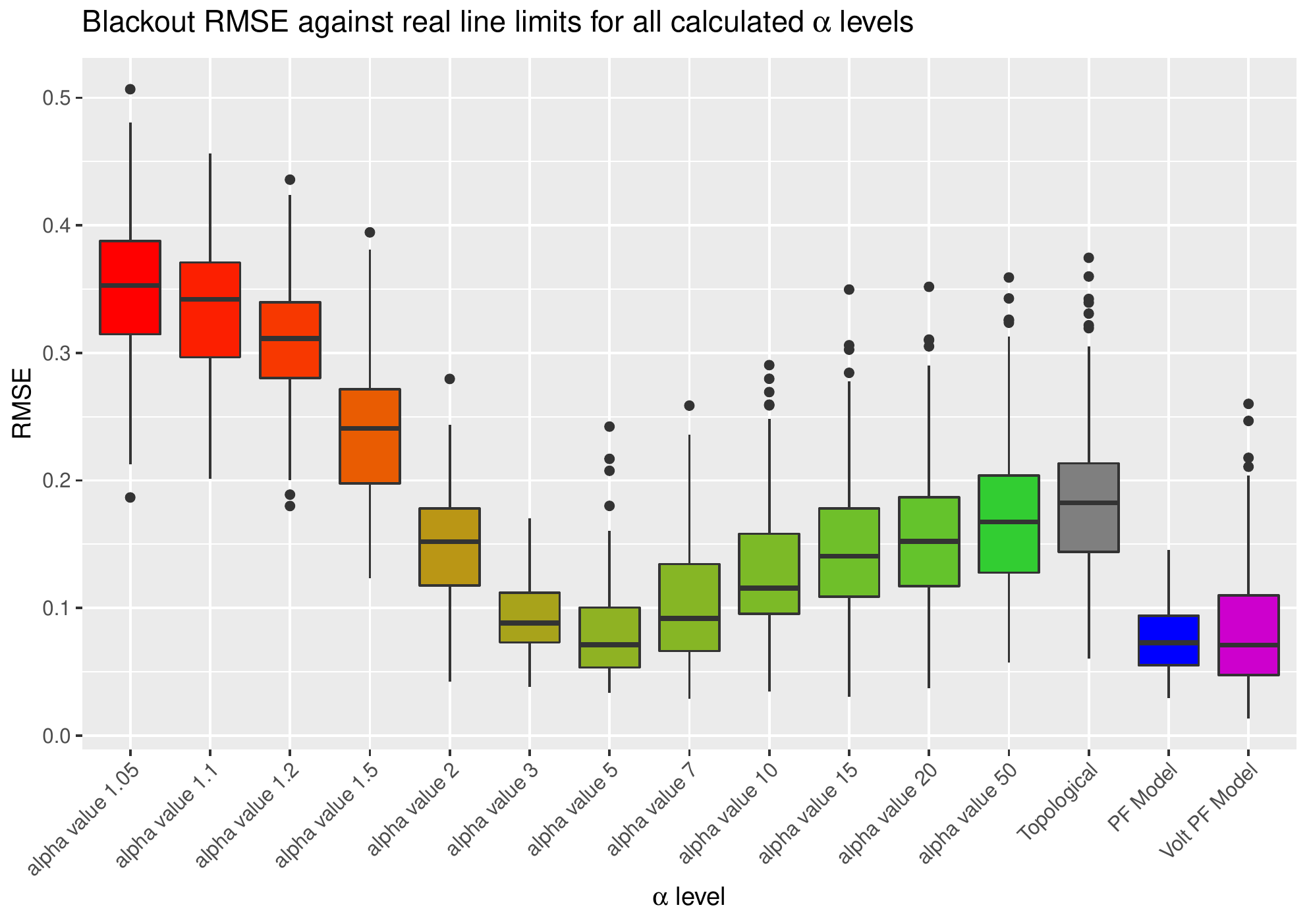}
    \caption{The PF model has the lowest RMSE with the real limits. However, the Volt PF model and $\alpha = 5$ are close behind.}
    \label{fig:RMSEchange}
\end{figure}

Figure \ref{fig:RMSEchange} shows that, despite its poor ability to accurately measure line limits, the PF model is the most accurate method of measuring blackout size, with an RSME of 0.077. The Volt PF model performs slightly worse than the $\alpha = 5$ (the system mean) in terms of blackout size. These models have an RMSE of 0.085 and 0.082 respectively.

For the giant component, the Volt PF model is the best performer with an RMSE error of 0.087. The Volt PF model is followed by $\alpha = 7$, $\alpha = 10$ which have RMSE scores of 0.103 and 0.104 respectively. The PF model comes fourth with an RMSE score 0.117. 

The error relationship between blackout size and optimum $\alpha$ is intuitive as it matches the average $\alpha$ of the real line limits. The reason for the giant-component optimum $\alpha$ is less apparent and may be linked to the topological structure. The linear models perform well in both cases.

\begin{figure}
    \centering
    \includegraphics{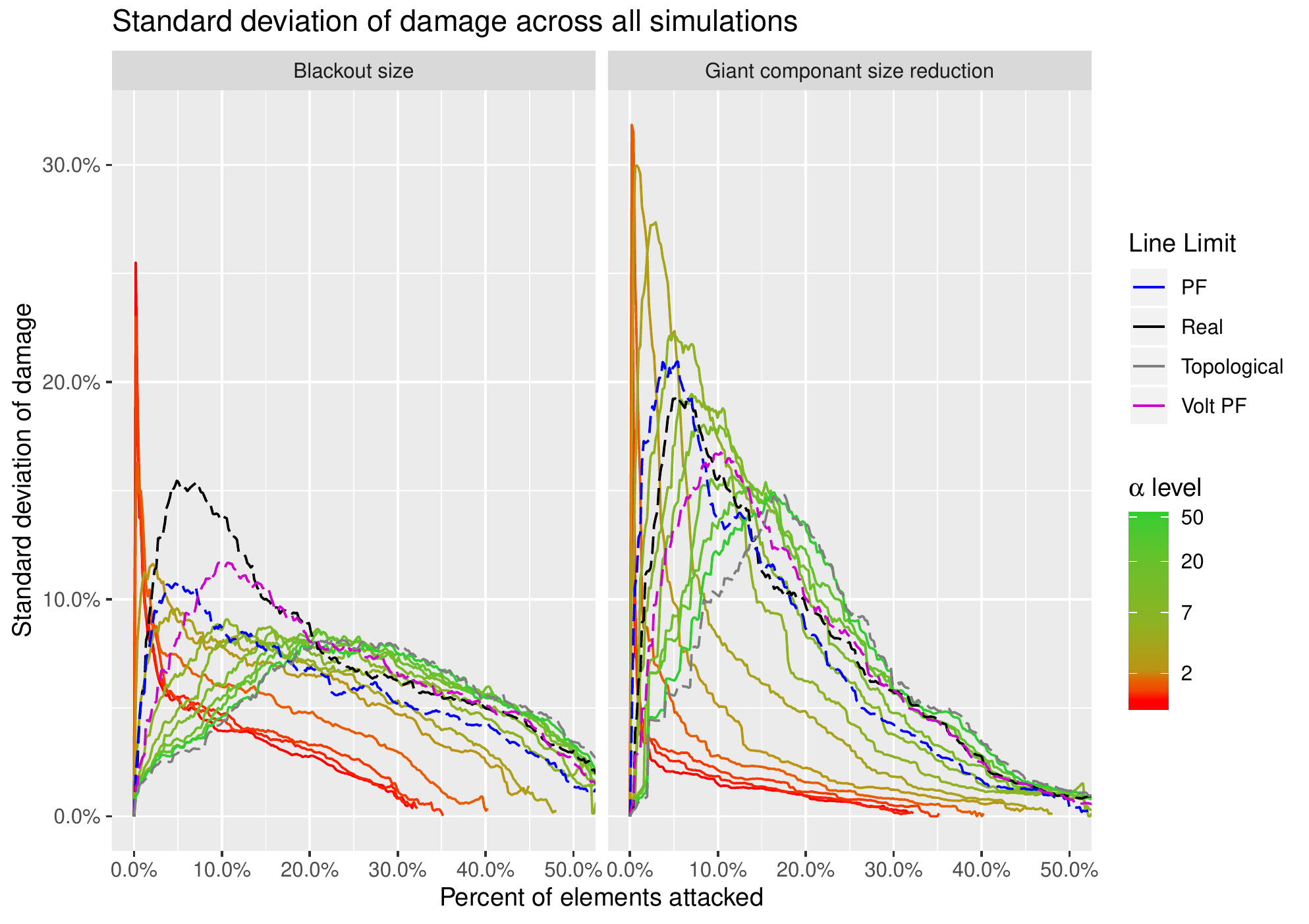}
    \caption{The standard deviation of the attack damage across the 100 simulations.}
    \label{fig:SDDamage}
\end{figure}

Next, we compare the standard deviation of damage across all 100 simulations. The PF Volt model has the best fit for the standard deviation of the blackout size, with an RMSE of 0.023, followed by $\alpha = 10$ (0.042). The PF model is in ninth place with an RMSE of 0.209. The accuracy of the PF Volt model comes from its ability to represent the hump of the real limits shown in Figure \ref{fig:SDDamage}, something no value of $\alpha$ can replicate.

When looking at the RMSE of the standard deviation of the damage of the giant component, we see that the Volt PF model and $\alpha = 10$, have similar score with 0.018 and and 0.021 respectively. The PF model is in eighth place with a score of 0.212. 

\begin{figure}
    \centering
    \includegraphics{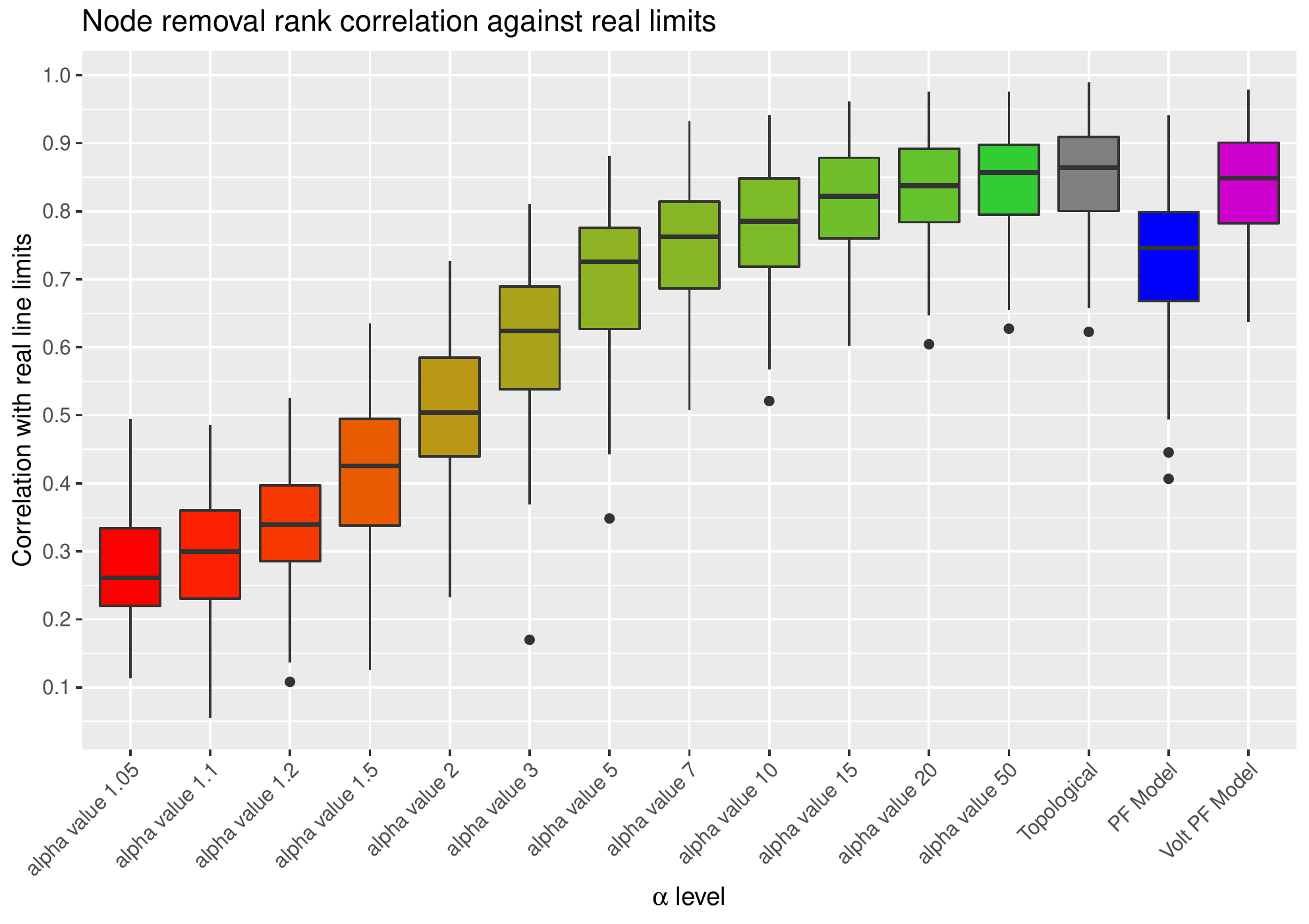}
    \caption{The Volt PF model has the highest correlation. Correlation of edge removal increases as $\alpha$ tends to infinity, which is identical to the topological analysis.}
    \label{fig:DamCor}
\end{figure}

We compare the correlation of the order of node loss between each artificial line limit and the real line limits. The correlation across all 100 simulations is shown in Figure \ref{fig:DamCor}. The figure shows that, as $\alpha$ increases, the node-loss-order similarity increases following a logarithmic growth curve with the topological analysis most similar to the real line limits.
The topological analysis has a mean correlation of 0.853, more than three times as high as $\alpha = 1.05$, which has a mean of 0.277. 
The $\alpha = 50$ model narrowly outperforms the Volt PF model, which have correlation scores of 0.847 and 0.842 respectively. The PF model has a significantly lower correlation score of 0.730.

\begin{figure}
    \centering
    \includegraphics{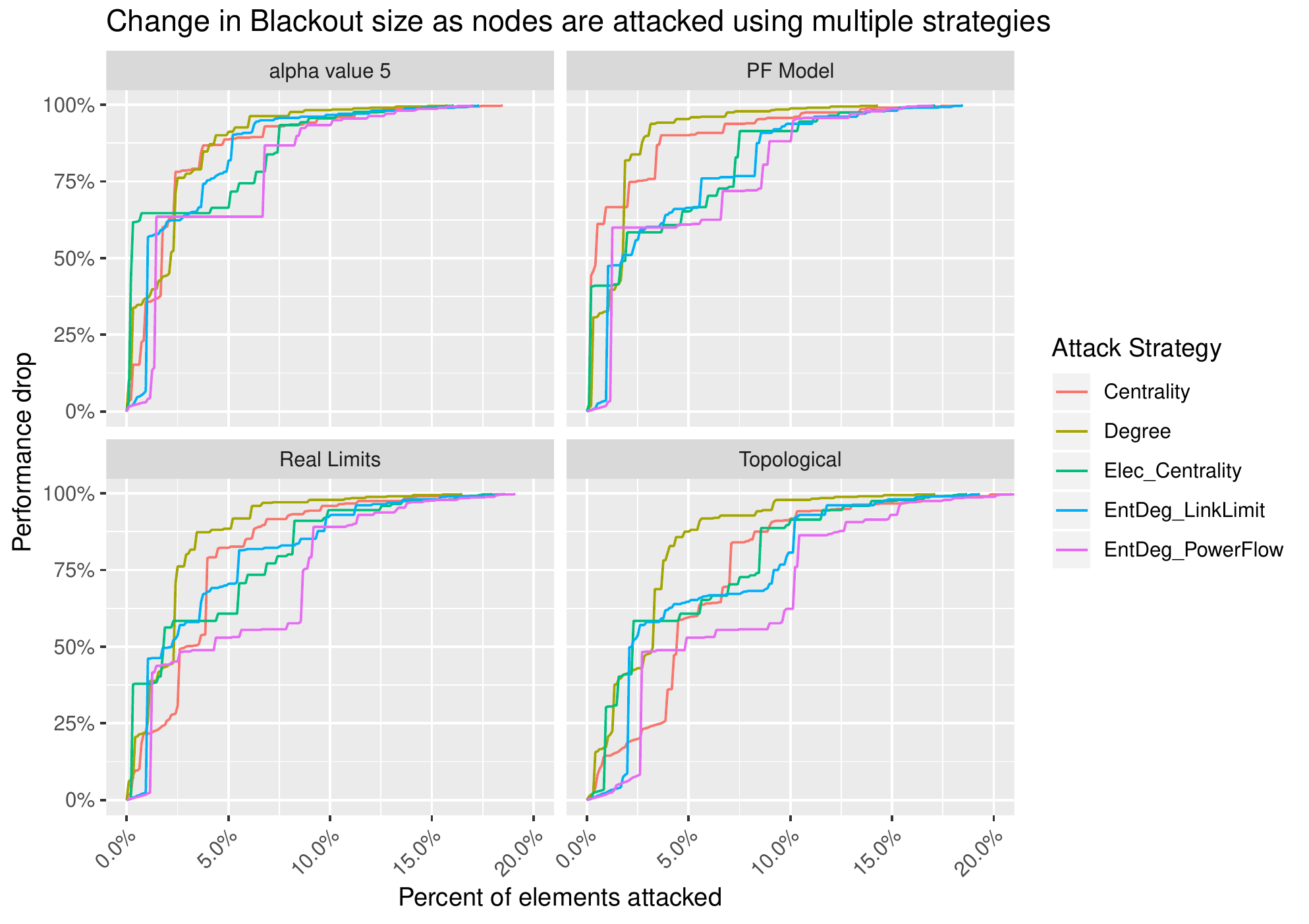}
    \caption{The figure shows the damage done to the power grid by each attack strategy by percentage of the grid attacked. As can be seen, strategy rank can change as a higher percentage of nodes is removed.}
    \label{fig:AttackStratDiff}
\end{figure}

\begin{figure}
    \centering
    \includegraphics{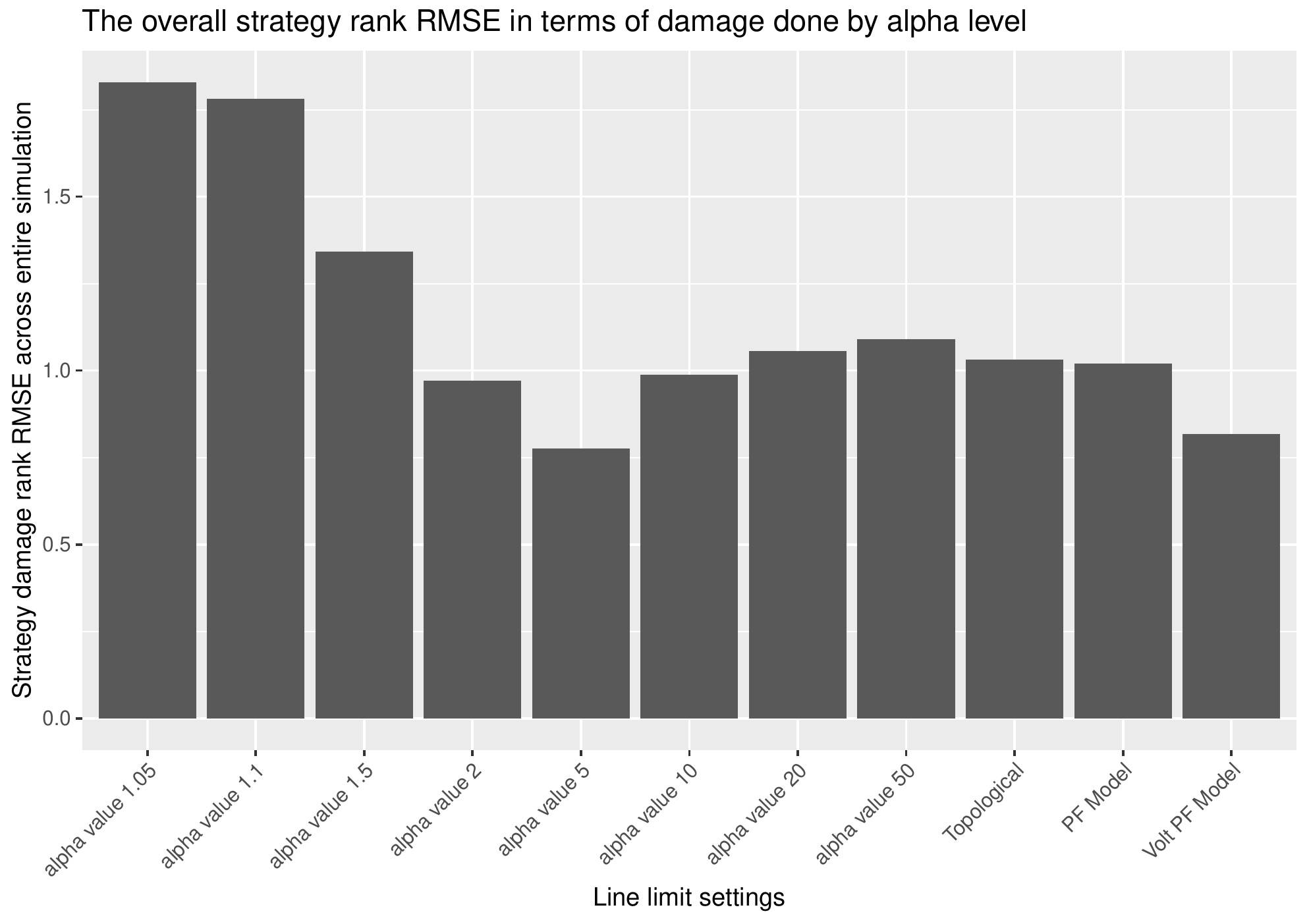}
    \caption{The $\alpha$ values equal to or lower than two are particularly poor at ranking attack strategies.}
    \label{fig:AttackStratRankPerf}
\end{figure}

Figure \ref{fig:AttackStratDiff} shows plots of the damage caused by the five attack strategies in four of the ten line-limit scenarios. As can be seen, the rankings of the attack strategies change as the number of nodes removed increases.
We analyse how well each line-limit method reflects the relative performance of the attack strategies. We find that that the PF Volt model (RMSE 0.818) is outperformed by $\alpha = 5$ (0.776), while the PF model (1.02) came fifth, narrowly beaten by $\alpha = 2$ (0.971) and $\alpha = 10$ (0.987) (see Figure \ref{fig:AttackStratRankPerf}). Values of $\alpha$ of 1.5 or less have a considerably greater error in ranking different attack strategies. 

That the low $\alpha$ values are much worse at ranking the performance of attack strategies is important as the majority of PL papers use $\alpha$ values of less than five. One reason for this may be that the cascade size is bigger with lower $\alpha$ values.

\begin{figure}
    \centering
    \includegraphics{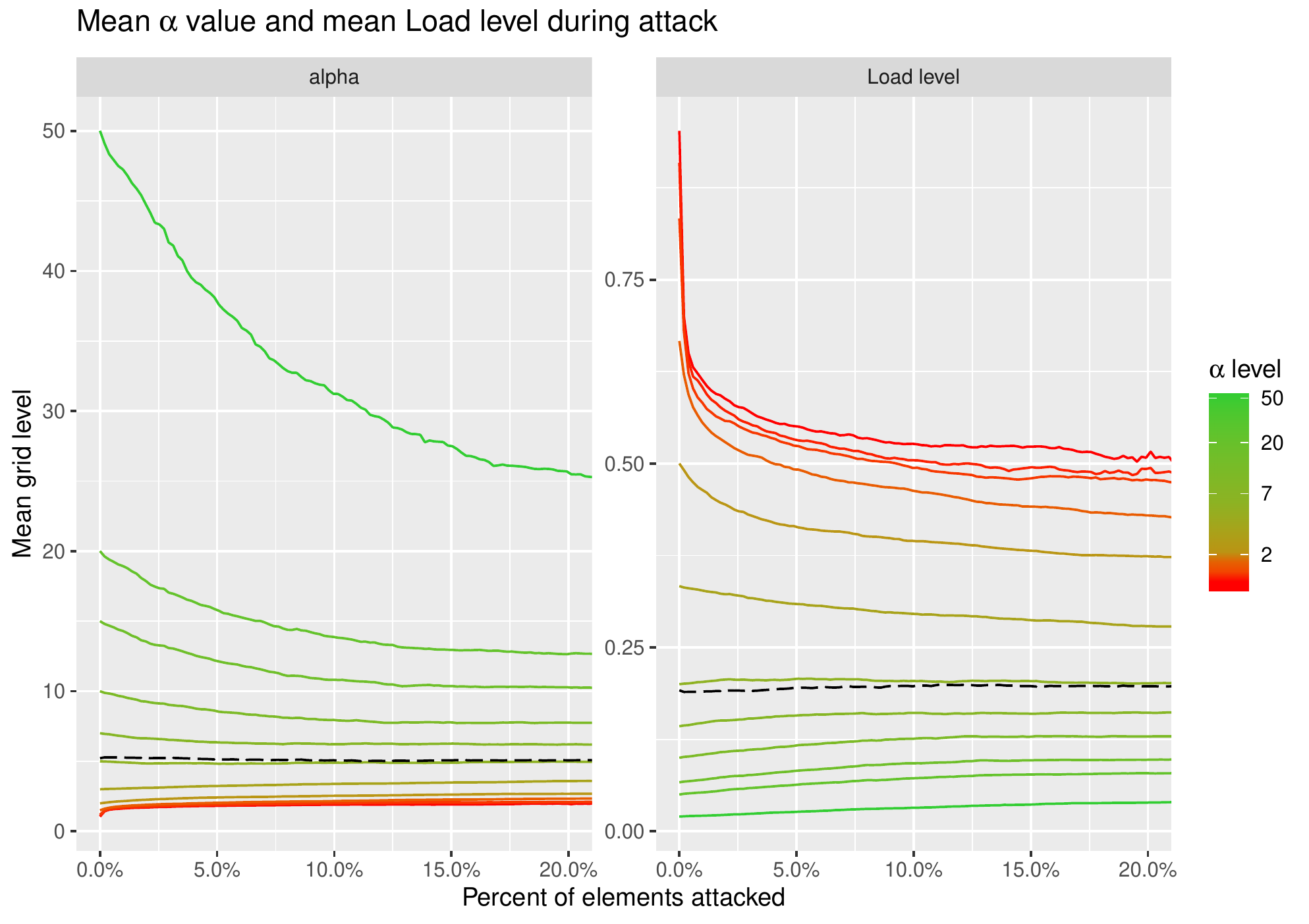}
    \caption{When the PL is a long way from the true system mean, a large drop is observed in $\alpha$ or $\frac{1}{\alpha}$ as the attack develops.}
    \label{fig:AlphaChange}
\end{figure}

\begin{figure}
    \centering
    \includegraphics{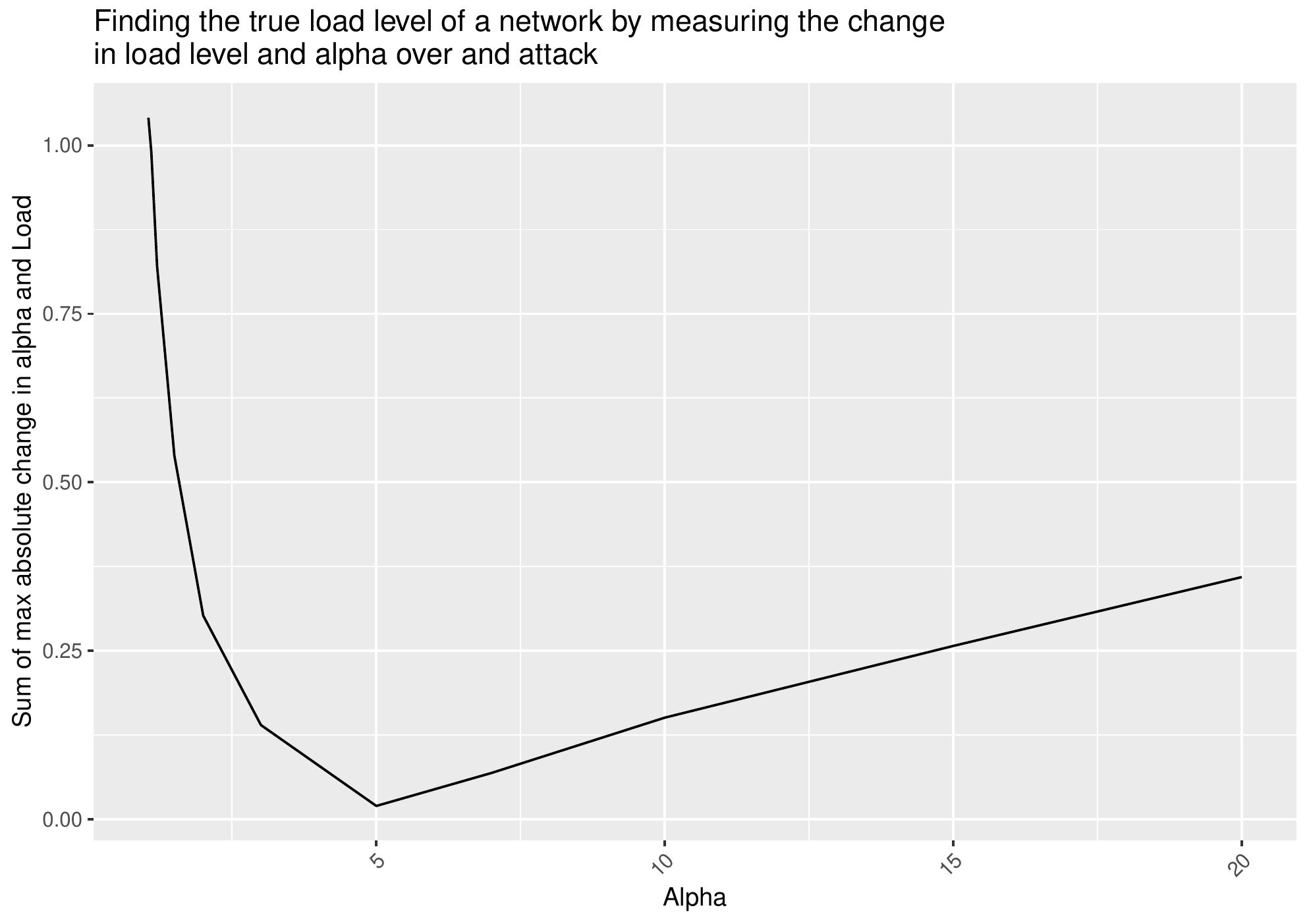}
    \caption{By summing the absolute maximum difference in values for $\alpha$ and load level, we find the minimum value is the true load level.}
    \label{fig:TrueAlpha}
\end{figure}

An interesting question is whether the network grid loading can provide any information about the system's true loading. Figure \ref{fig:AlphaChange} shows a large drop in $\alpha$ when the initial $\alpha$ of the system is much higher than the true system mean $\alpha$ of five. In contrast, the load level ($\frac{1}{\alpha}$) shows a large drop when the initial $\alpha$ of the system is much lower than the true system $\alpha$. When the initial $\alpha$ is set to the true system $\alpha$, the sum of the change in $\alpha$ and load level, $\Delta \text{Total} = \Delta \alpha + \Delta \frac{1}{\alpha}$, from initialisation is minimised (see Figure \ref{fig:TrueAlpha}).

\section{Discussion}

This paper has found that using linear models to estimate power-grid line limits gives consistently high performance. The Volt PF model performs very well across all tests whilst the PF performs better than most $\alpha$ values but is not outstanding. 
We find that the optimum $\alpha$ level is not consistent and changes depending on what is being measured. Although the system mean $\alpha =5$ performed well, sometimes other $\alpha$ values such as 10 or topological provided the best performance in specific analyses. Results based on lower values of $\alpha$ may may imply that the risk of cascading blackouts in some networks is greater than it actually is and give a false indication of the robustness of power networks to cascading failure. 

An interesting finding was that the ability of the linear model to measure line limits does not reflect the ability of the linear model to accurately represent power-grid collapse behaviour. Although the Volt PF model is much more accurate at estimating line limits, the PF is more accurate at modelling Blackout size during grid collapse. This may be because the PF model makes large errors on edges that do not overload and smaller errors on edges that do. If a line is unlikely to overload, the limit is irrelevant. A complicating factor is that the PF model has relatively poor correlation with the order in which the nodes are lost. This result is probably related to the overloaded lines, of which the PF model has more on average than the Volt PF model or the real line limits. Further research is needed to explore the relationship between modelling line limits and modelling collapse damage, especially regarding the importance of accurately identifying lines that may overload.

One of the major challenges of analysing power grids is that reasonable loading levels are very difficult to identify. In this paper, we found some evidence that, by analysing the change in loading under attack, we can identify the true value of the system $\alpha$. If true, the link between topology and $\alpha$ could be that the system has evolved as a transport mechanism for load and generation that are fixed geospatially and constrained in terms of MW range. These findings are preliminary. To provide firmer evidence, system attacks need to be simulated using a variety of different mean $\alpha$ and load/demand profiles.

\section{Conclusion}
The title of this paper suggests that it is unwise to use artificial line limits when studying the collapse behaviour of power grids under attack. In fact, given the importance of artificial line limits in this field, we merely propose that researchers consider carefully how they set these artificial line limits. This is because as we have shown, artificial line limits far from reality can produce results which themselves are far from reality and do not represent real grid behaviour. The goal of this paper has been to highlight the difficulties posed by artificial line limits and indicate some approaches which may lead to more accurate line limit modelling, and as a result solutions to cascading failures that are more applicable to real power networks.

\section*{acknowledgments}
This work was funded by EPRSC award 1828879

\newpage
\appendix

\section{Defining the simulation using the PEARL framework}  
\label{sect:PEARL}
Understanding a simulation's parameters is essential to understanding how a simulation works. A lack of data on the relative behaviour of different simulation methods makes it difficult to compare the results of different studies. To better understand simulation output, simulation parameters need to be clearly defined. Table \ref{tab:SimuParamsABS} gives a taxonomy of the simulation parameters used to define the experiments in this study. The subsequent subsections define each of the parameters and also the options implemented in the R Power Grid Networking package \cite{bourne_powergridnetworking_2018}. The simulation parameters proposed are not meant to be exhaustive but do describe key areas where clarity regarding the simulation strategy will help with the interpretation of the results and the replication of the experiments. N.B.: The attack strategies are deployed within a simulation and are not a part of the parameters.

\begin{table}[ht]
\centering
\caption{Classifications}
\label{tab:SimuParamsABS}
\begin{tabular}{l|l}
Class        & Types                     \\ \hline
\textbf{P}hysics      & Cascading DC, Topological \\
\textbf{E}lement      & Node, Edge, Both        \\
\textbf{A}ttack Type  & Fixed, Flexible, Adaptive \\
\textbf{R}emoval Regime     & Sequential, Simultaneous, hybrid  \\
\textbf{L}oad Profile & No sampling, Random sampling,\\
& Timeseries sampling, Random load        
\end{tabular}
\end{table}

\subsection{Physics}

The physics model is usually the most well-defined part of the simulation as it is the rule structure under which nodes and edges will operate. The line-flow limits and balancing methods are included as part of the physics model. Although there are many different physics models, in this package we consider only two.

\begin{itemize}
    \item Cascading DC: This model uses DC-flow calculations to decide whether line limits have been exceeded. This method can lead to cascading failures as power is redistributed across the network.
    \item Topological: This is a purely topological view of the power grid in which cascades are not possible. Although there are no cascades, multiple nodes can be lost in a single attack when a sub component is created that does not have both load and generation. This process can be reffered to as `Islanding', as the nodes are stranded on an island without power.
    A topological analysis is equivalent to proportional loading when $\alpha = \infty$
\end{itemize}

\subsection{Element}
Are nodes, edges or both being targeted during this attack? Certain attack strategies, such as degree, only work on one element type, while others can be applied to both, e.g., Net-ability \cite{arianos_power_2009}.

\begin{itemize}
    \item Nodes: Attack strategies only consider nodes.
    \item Edges: Attack strategies only consider edges.
    \item Both: Attack strategies consider nodes and edges. Can only be used for certain metrics.
\end{itemize}

\subsection{Attack type}
\label{sect:attacktype}
The attack type defines how an attack strategy will be executed. There are two classes of attack type: single calculation and repeated calculation. For single calculation, the attack order is calculated once before the attack begins. The repeated-calculation type only finds the next node to be removed and is calculated again for every attack round. The different attack types are defined below.

\begin{itemize}
    \item Fixed: This single-calculation method produces a target vector of $n$ nodes for removal. If some of those nodes are lost before being targeted, then the total amount of nodes targeted for removal will be $k-f$, where $f$ is the total number of nodes lost due to cascades or islanding.
 
    \item Flexible: This single-calculation method produces a target vector of $k$ nodes for removal and then removes $n$. This attack type is different from the fixed method as it will always remove $n$ nodes as long as the graph still has nodes to remove. If $n=20$ and the 19th node is lost in a cascade, the 21st node will be added to the target list.

    \item Adaptive: This is a repeated calculation in which the node order is recalculated at every attack. This attack type allows the order of removal to change as an attack develops. In a degree-based attack, after a few node removal rounds some previously high-degree nodes may have lost several neighbours. In contrast, other loads may not have lost any neighbours and be relatively more important.
\end{itemize}

\subsection{Removal regime}

The removal regime describes how nodes or edges will be removed from the network. The regime has implications if the physics model allows cascades. However, the removal regime has less of an effect in non-cascading simulations.

\begin{itemize}
    \item Sequential: The nodes on the target list are removed sequentially. The resulting cascades (if applicable) are calculated, and the grid is stabilised before the next node is removed.
    \item Simultaneous: The nodes in the target list are removed at the same time. This only makes sense when $k \gg n$.
    \item Hybrid: The nodes are removed in small groups, but the groups are removed sequentially. As an example, 30 nodes will be removed; however, they will be removed in three groups of 10.
\end{itemize}

\subsection{Load profile}

In many simulations, samples are taken to find a representative final result. In such cases, how the samples are chosen needs to be defined.

\begin{itemize}
    \item No sampling: The most straightforward method. Only a single attack run is made for each strategy using a given load profile.
    \item Random sampling: Only useful for the random attack strategy. The attack is repeated multiple times, and the results averaged. In most other attack strategies, the results will be identical.
    \item Time-series sampling: When a time series of loads is available for a network, it is randomly sampled to produce a variety of load profiles. Attacks can then be performed across the multiple time points, and the results analysed.
    \item Random-load sampling: Several papers have some variant of randomly assigning loads and generation across the network and then simulating attack. This method can only be used when line limits are proportional to initial loading.
\end{itemize}

\printbibliography

\end{document}